\documentclass[final,1p,times]{elsarticle}

 \usepackage{graphicx}
 \usepackage{color}

\usepackage{amssymb}

\usepackage{lineno}
\usepackage{bm}

\journal{Solid State Communications}

\begin{document}

\def\gsim{\mathop {\vtop {\ialign {##\crcr 
$\hfil \displaystyle {>}\hfil $\crcr \noalign {\kern1pt \nointerlineskip } 
$\,\sim$ \crcr \noalign {\kern1pt}}}}\limits}
\def\lsim{\mathop {\vtop {\ialign {##\crcr 
$\hfil \displaystyle {<}\hfil $\crcr \noalign {\kern1pt \nointerlineskip } 
$\,\,\sim$ \crcr \noalign {\kern1pt}}}}\limits}

\begin{frontmatter}



\title{
Elastic Softening in Quantum Critical Yb-Al-Au Approximant Crystal and Quasicrystal
}


\author{Shinji Watanabe}

\address{Department of Basic Sciences, Kyushu Institute of Technology, Kitakyushu, Fukuoka  804-8550, Japan}

\begin{abstract}
The elastic property of quantum critical quasicrystal (QC) Yb$_{15}$Al$_{34}$Au$_{51}$ is analyzed theoretically on the basis of the approximant crystal (AC) Yb$_{14}$Al$_{35}$Au$_{51}$. 
By constructing the realistic effective model in the AC, we evaluate the 4f-5d Coulomb repulsion at Yb as $U_{fd}\approx 1.46$~eV realizing the quantum critical point (QCP) of the Yb-valence transition. The RPA analysis of the QCP shows that softening in elastic constants occurs remarkably for bulk modulus and longitudinal mode at low temperatures. 
Possible relevance of these results to the QC as well as the pressure-tuned AC is discussed. 
\end{abstract}

\begin{keyword}
Heavy-fermion quasicrystal \sep Quantum valence criticality \sep Elastic constant


\end{keyword}

\end{frontmatter}



\section{Introduction}

Quantum critical phenomena discovered in quasicrystal (QC) Yb$_{15}$Al$_{34}$Au$_{51}$ have attracted great interest in condensed matter physics~\cite{Deguchi,Watanuki}. 
The measured criticality as the susceptibility $\chi\sim T^{-0.5}$, the specific-heat coefficient $C/T\sim-\log{T}$, the resistivity $\rho\sim T$, and the NMR relaxation rate $(T_1T)^{-1}\sim T^{-0.5}$ has been explained by the theory of critical Yb-valence fluctuations (CVF)~\cite{WM2010,WM2016,WMssc}. 
The CVF theory~\cite{WM2013} also elucidates not only the robust criticality in the QC against pressure but also emergence of the same criticality in the 1/1 approximant crystal (AC) Yb$_{14}$Al$_{35}$Au$_{51}$ when pressure is tuned to $P=1.96$~GPa~\cite{Matsukawa2016}.

Recently, a sharp change in the Yb valence has been observed at $x=0$ and $y=0$ in the QC 
Yb$_{15}$(Al$_{1-x}$Ga$_{x}$)$_{34}$(Au$_{1-y}$Cu$_{y}$)$_{51}$~\cite{NKSato2019}. This directly confirms the prediction of the CVF theory~\cite{WM2018}, which evidences the crucial role of the Yb valence in the quantum criticality. 
Since the local environment around Yb is common in both the QC and AC and the CVF are local in nature~\cite{WM2010}, the analysis of the CVF based on the AC~\cite{WM2016,WM2018} is considered to capture the essence of the quantum critical phenomena in the QC as well as the pressure-tuned AC.

In this paper, we discuss the elastic property of the QC Yb$_{15}$Al$_{34}$Au$_{51}$ on the basis of the analysis of the quantum critical point (QCP) of the Yb-valence transition in the AC. 
By constructing the effective model in the AC, 
we will show that softening in the elastic constants occurs remarkably for bulk modulus and longitudinal mode at the QCP. 

\begin{figure}
\includegraphics[width=9cm]{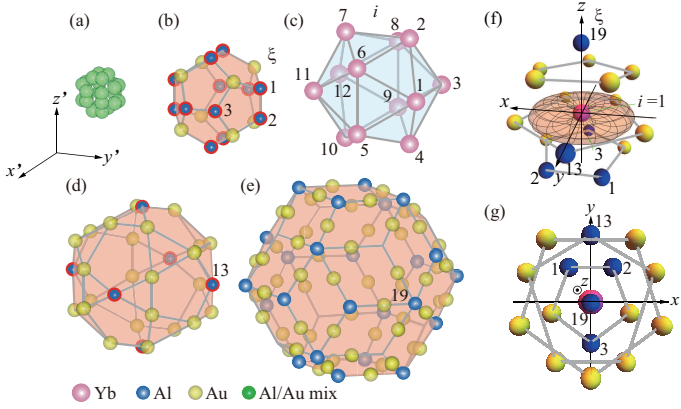}
\caption{(color online) 
%
Concentric shell structures of Tsai-type cluster in the Yb-Al-Au approximant crystal: (a) 1st shell, (b) 2nd shell (c) 3rd shell, (d) 4th shell, and (e) 5th shell. 
The number in (b), (d), (e), (f) and (g) indicates the $\xi$th Al site and number in (c) and (f) specifies the $i$th Yb site. 
(f) Local configuration around the Yb atom is shown in a side view. 
The pseudo fivefold axis, passing from the center of the icosahedron in (c) through the Yb atom, is illustrated as the $z$ axis. 
The square of absolute value of spherical part of the 4f wave function $\Psi^{4f}(\hat{\bm r})$ at the 1st Yb site is shown. 
(g) Top view around the 1st Yb site along the pseudo fivefold axis. 
}
\label{fig:YbAlAu}
\end{figure}

The QC Yb$_{15}$Al$_{34}$Au$_{51}$ consists of the Tsai-type cluster, 
which has concentric shell structures of the (a) 1st shell, (b) 2nd shell, (c) 3rd shell, (d) 4th shell, 
and (e) 5th shell shown in Fig.~\ref{fig:YbAlAu}.
The AC Yb$_{14}$Al$_{35}$Au$_{51}$ has the periodic structure of the body-centered cubic (bcc) 
where the Tsai-type cluster is located at each corner and center with the lattice constant 14.5~$\AA$~\cite{Ishimasa}. 

Let us start with the analysis of the crystalline 
electric field (CEF) in the QC and AC. 
Recently, the CEF in the AC Tm$_{3.5}$Sc$_{11}$Zn$_{85.5}$~\cite{Jazbec2016} and TbCd$_6$~\cite{Das2017} with the Tsai-type cluster has been observed in the specific-heat, susceptibility and neutron measurements. 
For Tm$^{3+} (4f^{12})$ and Tb$^{3+} (4f^8)$, the Hund's rule tells us that 
the total angular momentum in the CEF ground state is $J=6$. 
The measured CEF ground state in both systems is $|J=6,J_z=
\pm 6\rangle$. 
Since the wave function $\Phi^{4f}(\hat{\bm r})\equiv\langle\hat{\bm r}|J=6,J_z=
\pm 6\rangle$ is lying in the plane, on which the rare-earth ion is located, perpendicular to the fivefold axis [see Fig.~\ref{fig:YbAlAu}(f)], $\Phi^{4f}(\hat{\bm r})$ is interpreted to minimize the CEF energy in the hole picture of the $4f$ state so as to avoid the energy loss by Coulomb repulsion from surrounding positive ions (Zn$^{2+}$ or Cd$^{2+}$) on the upper and lower planes~\cite{noteCEF}. 

For Yb$^{3+} (4f^{13})$ in the Yb-Al-Au QC and AC, the total angular momentum in the CEF ground state is $J=7/2$. In the hole picture of the $4f$ state, to avoid the energy loss due to surrounding positive ions (Al$^{3+}$ and/or Au$^{1+}$) on the upper and lower planes, $|J=7/2, J_z=\pm 7/2\rangle$ is expected to minimize the CEF energy [see $\Psi^{4f}(\hat{\bm r})\equiv\langle\hat{\bm r}|J=7/2,J_z=
\pm 7/2\rangle$ shown in Fig.~\ref{fig:YbAlAu}(f)]. 
Furthermore, $\Psi^{4f}(\hat{\bm r})$ is also favorable from the hybridization picture of the CEF since 
$\Psi^{4f}(\hat{\bm r})$ can overlap considerably with the 3p wave function at the nearest-neighbor (N.N.) Al site [for example, the $\xi=13$th Al site for the $i=1$st Yb site in Fig.~\ref{fig:YbAlAu}(f)]. 

The importance of the 4f-3p hybridization was suggested by the measurements in the Yb-Ga-Au QC~\cite{Matsukawa}: 
It was reported that by replacing Al with Ga in the Yb-Al-Au QC, the quantum criticality disappears, suggesting that the Al-3p state contributes to the quantum critical state. 

As for the Yb 5d state, the Hund's rule tells us that the total angular momentum in the CEF ground state is $J=3/2$. 
The wave function $\Psi^{5d}(\hat{\bm r})\equiv\langle\hat{\bm r}|J=3/2, J_{z}=\pm 3/2\rangle$ is lying in the plane with the Yb atom perpendicular to the pseudo fivefold axis, which earns the considerable hybridization with the 3p state at the N.N. Al site [see Fig.~\ref{fig:YbAlAu}(f)]. Thus we employ this state as the CEF ground state.

In the QC Yb$_{15}$Al$_{34}$Au$_{51}$, 
from the entropy analysis by integrating the specific heat by temperature, the first excited CEF energy is at least larger than $\Delta=20$~K~\cite{Deguchi_PC}. We focus on the elastic property of the QC and AC at low temperatures for $T\ll\Delta$ in this study. 

\section{Construction of the model}

On the basis of the considerations noted above, 
we construct the effective model for the low-energy electronic states in the AC, which consists of Yb 4f and 5d orbitals and Al 3p orbitals in the hole picture as 
\begin{eqnarray}
H=H_p+H_f+H_d+H_{U_{fd}}+H_{fp}+H_{dp}.
\label{eq:Hamil}
\end{eqnarray}
In, Eq.~(\ref{eq:Hamil}), 
the 3p part is given by 
\begin{eqnarray}
H_{p}=\sum_{\langle j\xi, j'\xi'\rangle}\sum_{\sigma=\uparrow, \downarrow}\sum_{m, m'=z,\pm}
t^{pp}_{j\xi m, j'\xi' m'}p^{\dagger}_{j\xi m\sigma}p_{j'\xi' m'\sigma},  
\label{eq:Hp}
\end{eqnarray}
where the $p_{j\xi m\sigma} (p^{\dagger}_{j\xi m\sigma})$ operators annihilate (create) 3p holes at the $\xi$th Al site in the $j$th unit cell with $m=p_z, p_{\pm}$, 
where $p_{\pm}$ is defined as $p_{\pm}\equiv (p_x\pm i p_{y})/\sqrt{2}$, and spin $\sigma=\uparrow, \downarrow$. 
The energy levels of the $m=z, \pm$ states are set to be the same for simplicity of analysis, which is taken as the origin of the energy. 
Here, $\langle j\xi, j'\xi'\rangle$ denotes the pair of the Al sites, where 
we set the transfer integral $t^{pp}_{j\xi m, j'\xi' m'}$ 
for up to the 5th N.N. Al sites. 
Since the existence ratio of the atoms in the 1st shell is small (Al/Au: 7.8\%/8.9\%) [see Fig.~\ref{fig:YbAlAu}(a)]~\cite{Ishimasa}, here we consider the 2nd-5th shells. As a first step of analysis, the Al/Au mixed sites shown in the sites framed in red in Figs.~\ref{fig:YbAlAu}(b) and \ref{fig:YbAlAu}(d) are assumed to be all occupied by Al. There are 48 Al sites in the bcc unit cell of the AC.

The 4f part is given by 
\begin{eqnarray}
H_{f}=\sum_{j=1}^{N_L}\sum_{i=1}^{24}\left[
\varepsilon_{f}\sum_{\eta=\pm}n^{f}_{ji\eta}
+Un^{f}_{ji+}n^{f}_{ji-}
\right]
\label{eq:Hf}
\end{eqnarray}
with $N_L$ being the number of the unit cell and $n^{f}_{ji\eta}\equiv f^{\dagger}_{ji\eta}f_{ji\eta}$, where the $f_{ji\eta}$ $(f^{\dagger}_{ji\eta})$ operators annihilate (create) 4f holes with the Kramers state $\eta$ and the CEF ground energy $\varepsilon_{f}$ at the $i$th Yb site in the $j$th unit cell.

The 5d part is given by 
\begin{eqnarray}
H_{d}=\varepsilon_{d}\sum_{j=1}^{N_L}\sum_{i=1}^{24}\sum_{\eta=\pm}n^{d}_{ji\eta}
+\sum_{\langle ji, j'i'\rangle}\sum_{\eta, \eta'=\pm}t^{dd}_{ji\eta, j'i'\eta'}d^{\dagger}_{ji\eta}d_{j'i'\eta'}
\end{eqnarray}
with $n^{d}_{ji\eta}\equiv d^{\dagger}_{ji\eta}d_{ji\eta}$, where $d_{ji\eta}$ $(d^{\dagger}_{ji\eta})$ operators annihilate (create) 5d holes with the Kramers state $\eta$ and the CEF ground energy $\varepsilon_{d}$ at the $i$th 
Yb site in the $j$th unit cell. 
Here, $\langle ji, j'i'\rangle$ denotes the pair of the N.N. Yb sites [see Fig.~\ref{fig:YbAlAu}(c)] and 
$t^{dd}_{ji\eta, j'i'\eta'}$ is the transfer integral between the 5d states. 

The 4f-5d interaction is given by 
\begin{eqnarray}
H_{U_{fd}}=U_{fd}\sum_{j=1}^{N_L}\sum_{i=1}^{24}\sum_{\eta=\pm}\sum_{\eta'=\pm}n^{f}_{ji\eta}n^{d}_{ji\eta'}, 
\label{eq:H_Ufd}
\end{eqnarray}
where $U_{fd}$ is the onsite Coulomb repulsion between the 4f and 5d holes at the Yb site.

The 4f-3p hybridization is given by 
\begin{eqnarray}
H_{fp}=\sum_{\langle ji, j'\xi\rangle}\sum_{\eta=\pm}\sum_{\sigma=\uparrow, \downarrow}\left(
V_{ji\eta, j'\xi\sigma}^{fp}f^{\dagger}_{ji\eta}p_{j'\xi\sigma}+h.c.
\right), 
\label{eq:Hfp}
\end{eqnarray}
where $\langle ji, j'\xi\rangle$ denotes the pairs between the $i$th Yb site in the $j$th unit cell and the $\xi$th Al site in the $j'$th unit cell. 
We consider up to the 4th N.N. Yb-Al sites [for example, the pairs between the $i=1$st Yb site and the $\xi=1, 2, 3, 13$ and $19$th Al sites as shown in Fig.~\ref{fig:YbAlAu}(f)]. 

The 5d-3p hybridization is
\begin{eqnarray}
H_{dp}=\sum_{\langle ji, j'\xi\rangle}\sum_{\eta=\pm}\sum_{\sigma=\uparrow, \downarrow}\left(
V_{ji\eta, j'\xi\sigma}^{dp}d^{\dagger}_{ji\eta}p_{j'\xi\sigma}+h.c.
\right), 
\label{eq:Hdp}
\end{eqnarray}
where $\langle ji, j'\xi\rangle$ denotes the same Yb-Al pairs as noted in 
$H_{fp}$. 

In the calculations of the transfer integrals $t^{pp}_{j\xi m, j'\xi' m'}$, $t^{dd}_{ji\eta, j'i'\eta'}$ and hybridizations $V_{ji\eta, j'\xi\sigma}^{fp}$, $V_{ji\eta, j'\xi\sigma}^{dp}$, it is necessary to input the Slater-Koster parameters~\cite{Slater,Takegahara}. 
Here we use the general relations $(pp\pi)=(pp\sigma)/2$, $(pd\pi)=-(pd\sigma)/\sqrt{3}$, $(pf\pi)=-(pf\sigma)/\sqrt{3}$, $(dd\pi)=-2(dd\sigma)/3$, and $(dd\delta)=(dd\sigma)/6$, which were shown to hold in the linear muffin-tin orbital (LMTO) method~\cite{Andersen1977}. 
In this study, we set $(pp\sigma)=-1$ as the energy unit and set $(dd\sigma)=0.4$, $(pf\sigma)=-0.3$, and $(pd\sigma)=0.6$ in the hole picture as typical values. 
The distance dependences of transfer integrals and hybridizations between the $l$ and $l'$ orbitals where $l$ is the orbital angular momentum are set so as to follow $\sim 1/r^{l+l'+1}$ 
with $r$ being the distance of the atoms, following the LMTO argument~\cite{Andersen1977}. 

%
\begin{table}
\begin{center}
\begin{tabular}{cccccc}
\hline
\multicolumn{1}{c}{$i$} &\multicolumn{1}{c}{$\alpha_i$} & \multicolumn{1}{c}{$\beta_i$} & 
\multicolumn{1}{c}{\hspace*{0.3cm}$i$} &\multicolumn{1}{c}{$\alpha_i$} & \multicolumn{1}{c}{$\beta_i$} 
\\
\hline
$1$ & $\phi$ & $\pi/2$ & \hspace*{0.3cm} $7^*$ & $0$ & $\theta$ \\
$2^*$ & $\pi$ & $\theta$ & \hspace*{0.3cm}$8^*$ & $\pi/2$ & $-\phi$ \\
$3$ & $\pi-\phi$ & $\pi/2$ & \hspace*{0.3cm}$9^*$ & $-\pi/2$ & $\pi-\phi$ \\
$4^*$ & $0$ & $-\pi+\theta$ & \hspace*{0.3cm}$10^*$ & $\pi$ & $-\pi+\theta$ \\
$5^*$ & $\pi/2$ & $\pi-\phi$ & \hspace*{0.3cm}$11$ & $-\phi$ & $\pi/2$ \\
$6^*$ & $-\pi/2$ & $-\phi$ & \hspace*{0.3cm}$12$ & $-\pi+\phi$ & $\pi/2$ \\
\hline
\end{tabular}
\end{center}
\caption{Euler angles for each Yb site [see Fig.~\ref{fig:YbAlAu}(c)]. $^*$ indicates the Yb site performed by the $zxz$ rotation and otherwise by the $zyz$ rotation.}
\label{tb:Euler}
\end{table}

To calculate $t^{dd}_{ji\eta, j'i'\eta'}$, $V_{ji\eta, j'\xi\sigma}^{fp}$, and $V_{ji\eta, j'\xi\sigma}^{dp}$, the 4f and 5d wave functions at each Yb site, $\Psi^{4f}(\hat{\bm r}_i)$ and $\Psi^{5d}(\hat{\bm r}_i)$, are necessary, which can be obtained as follows: 
In the beginning, suppose that the local $xyz$ coordinate [see Figs.~\ref{fig:YbAlAu}(f) and \ref{fig:YbAlAu}(g)] coincides with the $x'y'z'$ coordinate in Fig.~\ref{fig:YbAlAu}(a). 
In the $x'y'z'$ coordinate, each axis is parallel to each side of the bcc unit cell. 
First, rotate the $x'y'z'$ coordinate around the $z'$ axis with the angle $\alpha$: $x'y'z' \to x''y''z'$. Second, rotate the coordinate around the $y''$~$(x'')$ axis with the angle $\beta$, giving rise to the local $xyz$ coordinate at each Yb site. 
These procedures are expressed in the so-called $zyz$~$(zxz)$ rotation. 
The transformation of the $|J, J_z\rangle$ state in the $zyz$ rotation is performed as 
\begin{eqnarray}
\hat{R}(\alpha,\beta,0)|JJ_{z}\rangle=\sum_{J_{z}'}D^{J}_{J_{z}'J_{z}}(\alpha,\beta,0)|JJ_{z}'\rangle, 
\label{eq:Rs2}
\end{eqnarray}
where $\hat{R}(\alpha,\beta,\gamma)=e^{-i\alpha J_{z}}e^{-i\beta J_{y}}e^{-i\gamma J_{z}}$ is the rotation operator for the Euler angles $\alpha$, $\beta$, and $\gamma$, and $D^{J}_{J_{z}'J_{z}}(\alpha,\beta,\gamma)$ is the Wigner's D matrix~\cite{Wigner}: 
\begin{eqnarray}
D^{J}_{J_{z}'J_{z}}(\alpha,\beta,\gamma)=e^{-i\alpha J_{z}'}d^{J}_{J_{z}'J_{z}}(\beta)e^{-i\gamma J_{z}}. 
\label{eq:WD}
\end{eqnarray}
Here, $d^{J}_{J_{z}'J_{z}}(\beta)$ is the Wigner's small d matrix~\cite{Wigner}, whose explicit form can be found in literatures~\cite{Messiah}. 
For the $zxz$ rotation, $\hat{R}$ is given by $\hat{R}(\alpha,\beta,\gamma)=e^{-i\alpha J_{z}}e^{-i\beta J_{x}}e^{-i\gamma J_{z}}$ and $d^{J}_{J_{z}'J_{z}}(\beta)$ in Eq.~(\ref{eq:WD}) should be replaced to $\tilde{d}^{J}_{J_{z}'J_{z}}(\beta)$, 
whose explicit form is given by 
\begin{eqnarray}
\tilde{d}^{J}_{J_{z}'J_{z}}(\beta)=\sum_{k}\frac{\sqrt{(J+J_{z})!(J-J_{z})!(J+J_{z}')!(J-J_{z}')!}}{(J+J_{z}-k)!k!(k-J_{z}+J_{z}')!(J-k-J_{z}')!}(-i)^{2k+J_{z}'-J_{z}}\left(\cos\frac{\beta}{2}\right)^{2J-2kJ_{z}-J_{z}'}\left(\sin\frac{\beta}{2}\right)^{2k+J_{z}'-J_{z}}. 
\label{eq:ddx}
\end{eqnarray}
Here the summation for $k$ is taken over $\max(0,J_{z}-J_{z}')\le k\le \min(J-J_{z}', J+J_{z})$. 

By applying the $zyz$ rotation to the $i=1, 3, 10$, and $12$th Yb site 
and the $zxz$ rotation to the $i=2, 4, 5, 6, 7, 8, 9$, and $11$th Yb site with $\alpha_i$, $\beta_i$, and $\gamma_i=0$ 
listed in Table~\ref{tb:Euler} with $\theta=31.97^{\circ}$ and $\phi=58.03^{\circ}$, the local wave function of $\Psi^{4f}(\hat{\bm r}_i)$ and $\Psi^{5d}(\hat{\bm r}_i)$ at each Yb site is obtained. 
In the rotated $xyz$ coordinate, the $z$ axis passing from the cluster center of the icosahedron [see Fig.~\ref{fig:YbAlAu}(c)] through the Yb atom is the quantization axis of the CEF, which is regarded as the pseudo fivefold axis [see Fig.~\ref{fig:YbAlAu}(g)].


\section{Results}

\subsection{QCP of valence transition}

To analyze electronic states in the AC Yb$_{14}$Al$_{35}$Au$_{51}$, 
we apply the slave-boson mean-field (MF) theory to Eq.~(\ref{eq:Hamil})~\cite{Read1983}. 
To describe the state for $U=\infty$ responsible for heavy electrons, we consider $f^{\dagger}_{ji\eta}b_{i\alpha}$ instead of $f^{\dagger}_{ji\eta}$ in Eq.~(\ref{eq:Hamil}) 
by introducing the slave-boson operator $b_{ji}$ to describe the $f^0$-hole state and requiring the constraint $\sum_{ji}\lambda_{ji}(\sum_{\eta=\pm}n^{f}_{ji\eta}+b^{\dagger}_{ji}b_{ji}-1)$ 
with $\lambda_{ji}$ being the Lagrange multiplier. 
For $H_{U_{fd}}$ in Eq.~(\ref{eq:H_Ufd}), we employ the MF decoupling as 
$U_{fd}n^{f}_{ji\eta}n^{d}_{ji\eta'}\approx U_{fd}\bar{n}^{f}_{i}n^{d}_{ji\eta'}+\bar{R}_{i}n^{f}_{ji\eta}-\bar{R}_{i}\bar{n}^{f}_{i}$, 
with $\bar{R}_{i}\equiv U_{fd}\bar{n}^{d}_{i}$
and $\bar{n}^{f (d)}_{i}\equiv\sum_{j\eta}\langle n^{f (d)}_{ji\eta}\rangle/N_{L}$. 
For simplicity we neglected the Fock term which has a finite value in the present locally-inversion symmetry broken system at Yb since previous study including the term did not change the main result~\cite{WM2018}.
Here we have assumed that MFs are uniform since we now focus on the paramagnetic-metal phase. Hence 
we have $\bar{b}_{i}=\langle b_{ji}\rangle$ and $\bar{\lambda}_{i}=\lambda_{ji}$. 
Then, by optimizing the Hamiltonian as $\partial\langle H\rangle/\partial\bar{b}_{i}=0$, $\partial\langle H\rangle/\partial\bar{\lambda}_{i}=0$,
and $\partial\langle H\rangle/\partial\bar{R}_{i}=0$, 
we obtain the set of the MF equations : 
\begin{eqnarray} 
\frac{1}{N_L}\sum_{{\bm k}\eta}\langle f_{{\bm k}i \eta}^{\dagger}f_{{\bm k}i \eta}\rangle+\bar{b}_{i}^2&=&1, 
\label{eq:MF1}
\\
\frac{1}{2N_{L}}\sum_{{\bm k}}\left[\sum_{\eta\xi m\sigma}V^{fp}_{{\bm k},\xi m\sigma, i{\eta}}\langle f^{\dagger}_{{\bm k}i{\eta}}p_{{\bm k}\xi m\sigma}\rangle
+h.c.\right]+\bar{\lambda}_{i}\bar{b}_{i}&=&0, 
\label{eq:MF2}
\\
\frac{1}{N_{L}}\sum_{{\bm k}\eta}\langle f^{\dagger}_{{\bm k}i{\eta}}f_{{\bm k}i{\eta}}\rangle
&=&
\bar{n}^{f}_{i}.
\label{eq:MF3}
\end{eqnarray}
The MF equations together with the equation for the half filling $\bar{n}=3$ are solved self-consistently, where the total filling $\bar{n}$ is defined by $\bar{n}\equiv\frac{1}{24}\sum_{i=1}^{24}(\bar{n}^{f}_{i}+\bar{n}^{d}_{i})+\frac{1}{48}\sum_{\xi=1}^{48}\bar{n}^{p}_{\xi}$ 
with $\bar{n}^{p}_{\xi}\equiv\sum_{{\bm k}m\sigma}\langle p^{\dagger}_{{\bm k}\xi m\sigma}p_{{\bm k}\xi m\sigma}\rangle/(3N_L)$. 

Below the results for $\varepsilon_{d}=-0.4$ in the $N_L=8^3$ system will be shown as a typical case. 
Figure~1(a) shows the $\varepsilon_{f}$ dependence of $\bar{n}^{f}$ $(=\bar{n}^{f}_i$ for $i=1-24)$ for the ground state. 
As $U_{fd}$ increases, the $\bar{n}^{f}$ change becomes sharp and the slope $-\partial\bar{n}^{f}/\partial\varepsilon_{f}$ diverges at $\varepsilon_{f}=-0.5992$ for $U_{fd}=0.364$. For $U>0.364$, a jump in $\bar{n}_{f}$ appears, which indicates the first-order valence transition. From these results, the QCP of the valence transition is identified to be $(\varepsilon_f^{QCP}, U_{fd}^{QCP})\approx (-0.5992, 0.364)$. 
At the QCP, $\bar{n}^{f}=0.69$ is realized, which is favorably compared with the intermediate valence Yb$^{+2.67}$ observed in the QC Yb$_{15}$Al$_{34}$Au$_{51}$~\cite{Watanuki}. 
The mass renormalization factor is obtained as $z_i\equiv \bar{b}_i^2\approx 0.30$ at the QCP, which reflects the quasiparticle state with the intermediate valence within the MF theory~\cite{note_C}.

Here we estimate the value of $(pp\sigma)$ as $(pp\sigma)\approx 4.01$~eV by inputting the N.N. Al-Al distance $d=2.48~\AA$ into the Harrison's formula $(pp\sigma)=\eta_{pp\sigma}\frac{\hbar^2}{md^2}$ with $\eta_{pp\sigma}=3.24$~\cite{WM2018,Harrison}. This gives $U_{fd}^{QCP}\approx 1.46$~eV in Eq.~(\ref{eq:H_Ufd}). 
Direct examination of the $U_{fd}$ value by experiments such as the partial-fluorescence-yield method in the X ray~\cite{Tonai2017} 
in the QC and also in the AC is an interesting future subject. 

\begin{figure}[t]
\includegraphics[width=7cm]{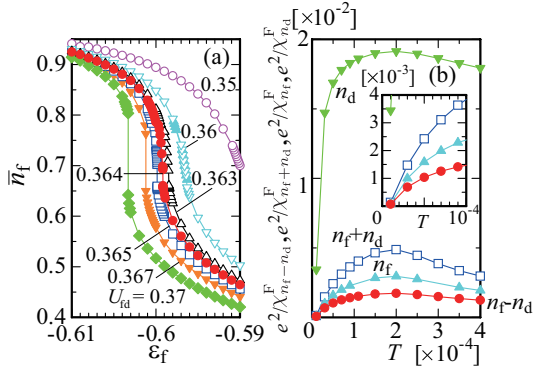}
\caption{(color online) 
(a) $\bar{n}_f$ vs. $\varepsilon_f$ for $U_{fd}=0.350$ (open circle), $0.360$ (open inverted triangle), $0.363$ (open triangle), $0.364$ (filled circle), $0.365$ (open square), $0.367$ (filled inverted triangle), and $0.370$ (filled diamond). 
(b) Temperature dependence of $e^2/\chi_{n_f-n_d}^{\rm F}$ (filled circle), $e^2/\chi_{n_f+n_d}^{\rm F}$ (open square), $e^2/\chi_{n_f}^{\rm F}$ (filled triangle), and $e^2/\chi_{n_d}^{\rm F}$ (filled inverted triangle) for  $(\varepsilon_f^{cRPA}, U_{fd}^{cRPA})$. Inset shows enlargement for $0\le T\le 10^{-4}$.
}
\label{fig:nfEf}
\end{figure}

It is noted that onsite 4f-5d hybridization occurs at Yb via the 4f-3p and 3p-5d hybridizations in Eq. (\ref{eq:Hamil}). This is nothing but the odd-parity CEF due to the local violation of the inversion symmetry at the Yb site by the fivefold configuration of the surrounding Au and Al sites [see Fig.~\ref{fig:YbAlAu}(g)].


Next, let us discuss the elastic property near the QCP. 
By using the random phase approximation (RPA) with respect to $U_{fd}$ as the corrections to the MF state, 
electric quadrupole fluctuations near the QCP can be analyzed. 
The RPA susceptibility is given by~\cite{THU2004} 
\begin{eqnarray}
& &\chi^{ii'}_{\ell_1\eta_1\ell_2\eta_2\ell_3\eta_3\ell_4\eta_4}({\bm q},\omega)
=\bar{\chi}^{ii'}_{\ell_1\eta_1\ell_2\eta_2\ell_3\eta_3\ell_4\eta_4}({\bm q},\omega)
\nonumber
\\
&-&\sum_{i''}\sum_{\tau\tau'}\sum_{m\ne m'}
\bar{\chi}^{ii''}_{\ell_1\eta_1\ell_2\eta_2m\tau m\tau}({\bm q},\omega)
U_{\rm fd}
\chi^{i'' i}_{m'\tau' m'\tau'\ell_3\eta_3\ell_4\eta_4}({\bm q},\omega)
\nonumber
\\
&+&\sum_{i''}\sum_{\tau\tau'}\sum_{m\ne m'}\bar{\chi}^{ii''}_{\ell_1\eta_1\ell_2\eta_2 m\tau m'\tau'}({\bm q},\omega)
U_{\rm fd}
\chi^{i'' i'}_{m\tau m'\tau'\ell_3\eta_3\ell_4\eta_4}({\bm q},\omega). 
\label{eq:RPA}
\end{eqnarray}
Here, the susceptibility is defined by 
\begin{eqnarray}
\chi^{ii'}_{\ell_1\eta_1\ell_2\eta_2\ell_3\eta_3\ell_4\eta_4}({\bm q},\omega)
\equiv\frac{i}{N_{
L
}
}\sum_{{\bm k}{\bm k'}}\int_{0}^{\infty}dte^{i\omega t}
\nonumber
\\
\times
\langle[c^{\dagger}_{{\bm k}i\ell_1\eta_1}(t)c_{{\bm k}+{\bm q}i\ell_2\eta_2}(t), c^{\dagger}_{{\bm k'}+{\bm q}i'\ell_4\eta_4}c_{{\bm k'}i'\ell_3\eta_3}]\rangle, 
\label{eq:chi_def}
\end{eqnarray}
where $\ell=1 (2)$ specifies the f (d) orbital and 
$\bar{\chi}^{ii''}_{\ell_1\eta_1\ell_2\eta_2\ell_3\eta_3 \ell_4\eta_4}({\bm q},\omega)$ denotes the susceptibility in the MF state.
Here the susceptibility is expressed in the retarded form, which is obtained by analytic continuation from the Matsubara form~\cite{THU2004}.

The valence transition is caused by the onsite Coulomb repulsion $U_{fd}$ which affects heavy quasiparticles realized by large $U (\gg U_{fd})$~{\cite{WIM2006}}. 
To describe this, we first obtained the QCP of the valence transition within the MF theory for $U\to\infty$ and then to take account of critical fluctuations caused by $U_{fd}$ beyond the MF theory, Eq.~(\ref{eq:RPA}) is employed. 
This approach is based on the physical picture discussed in the mode-mode-coupling theory of the CVF~\cite{WM2010}, which describes the quantum valence criticality observed in experiments~\cite{Deguchi, NKSato2019,Nakatsuji, Matsumoto,Kuga2018} correctly. The detail of the formulation was discussed in Ref~\cite{WM2019,W2020}. 
The RPA susceptibility in Eq.~(\ref{eq:RPA}) is expressed by the 
$384\times 384$ matrix $\hat{\chi}$, $\hat{\bar{\chi}}$ and $\hat{U}$ in the symmetrized form as 
\begin{eqnarray}
\hat{\chi}
=\hat{\bar{\chi}}^{1/2}\left(\hat{1}-\hat{\bar{\chi}}^{1/2}\hat{U}\hat{\bar{\chi}}^{1/2}\right)^{-1}\hat{\bar{\chi}}^{1/2},
\label{eq:RPA_chi}
\end{eqnarray}
where $\hat{\bar{\chi}}^{1/2}$ is the matrix satisfying $\hat{\bar{\chi}}=\hat{\bar{\chi}}^{1/2}\hat{\bar{\chi}}^{1/2}$ and $\hat{1}$ is the identity matrix
~\cite{W2020}. 
The interaction matrix $\hat{U}$ has the elements of $U_{\rm fd}$ for $\ell_1=\ell_3\ne\ell_2=\ell_4$, $\eta_1=\eta_3$, and $\eta_2=\eta_4$
and $-U_{\rm fd}$ for $\ell_1=\ell_2\ne\ell_3=\ell_4$, $\eta_1=\eta_2$, and $\eta_3=\eta_4$. 

The critical point in this RPA formalism is determined by 
\begin{eqnarray}
{\rm det}\left(\hat{1}-\hat{\bar{\chi}}^{1/2}\hat{U}\hat{\bar{\chi}}^{1/2}\right)=0. 
\label{eq:RPA_det0}
\end{eqnarray}
By using the MF states obtained in the calculation in Fig.~\ref{fig:nfEf}(a), we calculate $\hat{\bar{\chi}}$ on the basis of Eq.~(\ref{eq:chi_def}). 
Then, by solving Eq.~(\ref{eq:RPA_det0}), 
the critical point in this RPA formalism is obtained as $(\varepsilon_{f}^{c RPA}, U_{fd}^{c RPA})=(-0.5992, 01116)$. 
We confirmed that the maximum appears at ${\bm q}={\bf 0}$ in $\hat{\bar{\chi}}({\bm q},\omega=0)$ with weak ${\bm q}$ dependence for finite ${\bm q}$, as shown in Ref.~\cite{WM2016}. This is the reflection of the uniform CVF with the local nature as discussed in Ref.~\cite{WM2010}.

\subsection{Relative and total charge fluctuations}

At the valence QCP, the relative-charge fluctuations between the 4f and 5d orbitals, i.e., the CVF diverge, as shown in Fig.~\ref{fig:nfEf}(b). 
Here, we plot the temperature dependence of the uniform relative- and total-charge susceptibility $\chi^{\rm F}_{n_f\pm n_d}$ given by $\chi^{\rm F}_{n_f\pm n_d}=\lim_{{\bm q}\to{\bf 0}}\chi^{\rm F}_{n_f\pm n_d}({\bm q},\omega=0)$ with $\chi^{\rm F}({\bm q},\omega)=\sum_{ii'}\chi^{ii'}_{n_f\pm n_d}({\bm q},\omega)$ for $(\varepsilon_f^{cRPA}, U_{fd}^{cRPA})$. Here, $ \chi^{ii'}_{n_{f}\pm n_{d}}({\bm q},\omega) $ is defined by 
\begin{eqnarray}
\chi^{ii'}_{n_{f}\pm n_{d}}({\bm q},\omega)&=&\frac{ie^2}{N_L}\int_{0}^{\infty}dte^{i\omega t}\langle[\delta n^{f}_{\bm{q}i}(t)\pm \delta n^{d}_{\bm{q}i}(t), \delta n^{f}_{-\bm{q}i'}(0)\pm \delta n^{d}_{-\bm{q}i'}(0) ]\rangle, 
\label{eq:chi_nf_nd1}
\end{eqnarray}
with $+(-)$ denoting the total (relative) charge susceptibility and $e$ being an elementary charge $(e>0)$. 
Here $\delta\hat{\cal O}$ is defined as $\delta\hat{\cal O}\equiv\hat{\cal O}-\langle\hat{\cal O}\rangle$ 
and $n^{f (d)}_{{\bm q}i}$ is given by
$n^{f (d)}_{{\bm q}i}=\sum_{j}e^{-i{\bm q}\cdot\bm{r}_j}n^{f (d)}_{ji}$. 
The divergence of $\chi^{\rm F}_{n_f-n_d}$ for $T\to 0$ is confirmed by the eigenvalue analysis of the kernel $\hat{\bar{\chi}^{1/2}\hat{U}\hat{\bar{\chi}}^{1/2}}$, as shown in Fig.~\ref{fig:chi}(a). For $T\to 0$, the largest eigenvalue is identified to be that of the relative-charge fluctuation, which approaches 1. This satisfies Eq.~(\ref{eq:RPA_det0}) at $T=0$, giving rise to the valence QCP within the RPA.

Although the total charge fluctuation is also enhanced on cooling in Fig.~\ref{fig:nfEf}(b), it turns out that $\chi^{\rm F}_{n_f+n_d}$ does not diverge at $T=0$ from the analysis of the eigenvalue of the kernel, which does not approach 1 as shown in Fig.~\ref{fig:chi}(a). 
The difference between $\chi_{n_f-n_d}^{\rm F}$ and $\chi_{n_f+n_d}^{\rm F}$ comes from the negative value of the interorbital term in the right hand side (r.h.s.) of Eq. (\ref{eq:chi_nf_nd1}), as a consequence of $U_{fd}$, which forces the f and d occupancies to compete (see, for detail, Ref.~\cite{W2020}).

\subsection{Electric quadrupole fluctuations}

\begin{figure}[t]
\includegraphics[width=7cm]{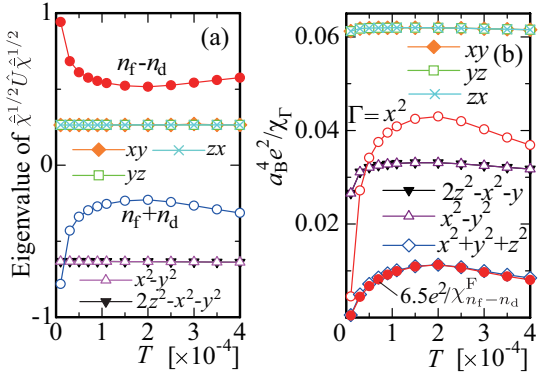}
\caption{(color online) 
(a) Temperature dependence of eigenvalues of $\hat{\bar{\chi}}^{1/2}\hat{U}\hat{\bar{\chi}}^{1/2}$ for electric monopoles [relative charge (filled circle) and total charge (open circle)] and electric quadrupoles [$2z^2-x^2-y^2$ (filled inverted triangle), $x^2-y^2$ (open triangle), $xy$ (filled diamond), $yz$ (open square), and $zx$ (cross) ] at $(\varepsilon_f^{cRPA}, U_{fd}^{cRPA})$. 
(b) Temperature dependence of electric quadrupole susceptibility $a_{\rm B}^2e^2/\chi_{\Gamma}$ for $\Gamma=x^2+y^2+z^2$ (open diamond), $2z^2-x^2-y^2$ (filled inverted triangle), and $x^2-y^2$ (open triangle), $xy$ (filled diamond), $yz$ (open square), $zx$ (cross) and $x^2$ (open circle) for $(\varepsilon_{f}^{RPA}, U_{fd}^{RPA})$. The $T$ dependence of $6.5e^2/\chi^{\rm F}_{n_f-n_d}$ is also shown (filled circle). 
}
\label{fig:chi}
\end{figure}

Next, let us discuss the electric quadrupole fluctuation near the QCP of the valence transition. 
The electric quadrupole susceptibility is expressed as 
\begin{eqnarray}
\chi^{ii'}_{O_{\Gamma}}({\bm q},\omega)=\frac{i}{N}\int_{0}^{\infty}dte^{i\omega t}\langle[\delta\hat{O}^{\Gamma}_{\bm{q}i}(t), \delta\hat{O}^{\Gamma}_{\bm{q}i'}(0)^{\dagger}]\rangle, 
\label{eq:chi_E4}
\end{eqnarray}
where 
$\hat{O}^{\Gamma}_{{\bm q}i}$ is given by 
%
$\hat{O}^{\Gamma}_{{\bm q}i}=\sum_{j}e^{-i{\bm q}\cdot\bm{r}_j}\hat{O}_{ji}
^{
\Gamma
}
$ 
%
with the irreducible representation $\Gamma$. 
In the AC Yb$_{14}$Al$_{35}$Au$_{51}$, the crystal structure is
cubic (space group No. 204, $Im\bar{3}$, $T_h^5$)~\cite{Ishimasa}. 
Then, $\Gamma$ is given by $\Gamma=x^2+y^2+z^2$, $2z^2-x^2-y^2$, $x^2-y^2$, $xy$, $yz$, and $zx$. 
Here $\hat{O}^{\Gamma}_{ji}$ is the quadrupole operator 
\begin{eqnarray}
\hat{O}^{\Gamma}_{ji}=\sum_{\ell\ell'}\sum_{\eta\eta'}O^{\Gamma}_{i\ell\eta, i\ell'\eta'}c^{\dagger}_{ji\ell\eta}c_{ji\ell'\eta'}, 
\label{eq:Oi}
\end{eqnarray}
where $O^{\Gamma}_{i\ell\eta, i\ell'\eta'}$ is the form factor 
\begin{eqnarray}
O^{\Gamma}_{i\ell\eta, i\ell'\eta'}=
e\langle r^2\rangle_{\ell}\alpha_{\ell}
\langle i\ell\eta|\hat{O}_{\Gamma}|i\ell'\eta'\rangle 
\label{eq:Omat}
\end{eqnarray}
with  $\langle r^2 \rangle_{\ell}$ being the expectation value of radial part of 4f and 5d wavefunctions. Here, we set $\langle r^2 \rangle_{1}=0.1826$~\AA$^2$ following the result of Yb$^{+3}$~\cite{Freeman}. As for the 5d orbital at Yb, we set $\langle r^2 \rangle_{2}=0.3246$~\AA$^2$ so as to reproduce the eigenvector of the kernel $\hat{\bar{\chi}}^{1/2}\hat{U}\hat{\bar{\chi}}^{1/2}$ for the electric-quadrupole modes of $\Gamma=2z^2-x^2-y^2$, $x^2-y^2$, $xy$, $yz$, and $zx$, whose eigenvalues are shown in Fig.~\ref{fig:chi}(a). 
Note that the relative value of $\langle r^2\rangle_2/\langle r^2\rangle_1$ is important here and choice of the value of  $\langle r^2\rangle_1$ merely shifts the whole scale of $\chi_{\Gamma}$ 
as far as the ratio $\langle r^2\rangle_2/\langle r^2\rangle_1$ is maintained,
which does not affect the discussions below. 
In Eq.~(\ref{eq:Omat}), the Stevens factor is given by $\alpha_1=2/63$~\cite{Stevens} and $\alpha_2=-2/15$. 
Then, Eq. (\ref{eq:chi_E4}) leads to
\cite{Kontani2011}
\begin{eqnarray}
\chi^{ii'}_{O_{\Gamma}}({\bm q},\omega)=\sum_{\ell_1\ell_2\ell_3\ell_4}\sum_{\eta_1\eta_2\eta_3\eta_4}O^{\Gamma}_{i\ell_1\eta_1, i\ell_2\eta_2}\chi_{\ell_1\eta_1\ell_2\eta_2\ell_3\eta_3\ell_4\eta_4}^{ii'}({\bm q},\omega)O^{\Gamma}_{i'\ell_4\eta_4, i'\ell_3\eta_3}. 
\label{eq:chi_E4_2}
\end{eqnarray}
%

In Eq.~(\ref{eq:Omat}), the operators $\hat{O}_{\Gamma}$ are expressed by the total angular momentum 
in the laboratory coordinate 
as 
$\hat{O}_{x^2+y^2+z^2}=J_{x'}^2+J_{y'}^2+J_{z'}^2$, 
$\hat{O}_{2z^2-x^2-y^2}=
(
2J_{z'}^2-J_{x'}^2-J_{y'}^2
)/2
$, 
$\hat{O}_{x^2-y^2}=
\sqrt{3}
(
J_{x'}^2-J_{y'}^2
)/2
$, 
$\hat{O}_{xy}=
\sqrt{3}
(
J_{x'}J_{y'}+J_{y'}J_{x'}
)/2
$, $\hat{O}_{yz}=
\sqrt{3}
(
J_{y'}J_{z'}+J_{z'}J_{y'}
)/2
$, and 
$\hat{O}_{zx}=
\sqrt{3}
(
J_{z'}J_{x'}+J_{x'}J_{z'}
)/2
$~\cite{HYYK2018}.

The relation between 
the total angular momentum in the $x'y'z'$ coordinate and the one in the local $xyz$ coordinate is obtained by using $\hat{R}(\alpha_i,\beta_i,0)$ in Eq.~(\ref{eq:Rs2}) for the $i=1$, 3, 11, and 12th Yb site as 
%
$J_{x'}=J_{x}\cos\alpha_i\cos\beta_i-J_{y}\sin\alpha_i+J_{z}\cos\alpha_i\sin\beta_i$, 
$J_{y'}=J_{x}\sin\alpha_i\cos\beta_i+J_{y}\cos\alpha_i+J_{z}\sin\alpha_i\sin\beta_i$, and 
$J_{z'}=-J_{x}\sin\beta_i+J_{z}\cos\beta_i$.
%
By using these equations and $\Psi^{4f}(\hat{\bm r}_i)$ and $\Psi^{5d}(\hat{\bm r}_i)$, 
the form factors are calculated as 
$\langle i\ell\pm|\hat{O}_{x^2+y^2+z^2}|i\ell\pm\rangle=a_{\ell}(a_{\ell}+1)$, 
$\langle i\ell\pm|\hat{O}_{2z^2-x^2-y^2}|i\ell\pm\rangle=a_{\ell}(2-a_{\ell})(\sin^2\beta_i-2\cos^2\beta_i)
/2
$, 
$\langle i\ell\pm|\hat{O}_{x^2-y^2}|i\ell\pm\rangle=
\sqrt{3}
a_{\ell}(2-a_{\ell})(\sin^2\alpha_i-\cos^2\alpha_i)\sin^2\beta_i
/2
$, 
$\langle i\ell\pm|\hat{O}_{xy}|i\ell\pm\rangle=
\sqrt{3}
a_{\ell}(-1+2a_{\ell})\cos\alpha_{i}\sin\alpha_{i}\sin^2\beta_{i}
/2
$, 
$\langle i\ell\pm|\hat{O}_{yz}|i\ell\pm\rangle=
\sqrt{3}
a_{\ell}(-1+2a_{\ell})\sin\alpha_{i}\sin\beta_{i}\cos\beta_{i}
/2
$, and 
$\langle i\ell\pm|\hat{O}_{zx}|i\ell\pm\rangle=
\sqrt{3}
a_{\ell}(-1+2a_{\ell})\cos\alpha_{i}\sin\beta_{i}\cos\beta_{i}
/2
$, 
where $a_{\ell}$ is given by $a_{1}=7/2$ for the 4f state and $a_{2}=3/2$ for the 5d state, i.e., the total angular momentum. 

For the $i=2$, 4, 5, 6, 7, 8, 9, and 10th Yb site, we have 
%
$J_{x'}=J_x\cos\alpha_i-J_y\sin\alpha_i\cos\beta_i+J_z\sin\alpha_i\beta_i$,
$J_{y'}=J_x\sin\alpha_i+J_y\cos\alpha_i\cos\beta_i-J_z\cos\alpha_i\sin\beta_i$, and 
$J_{z'}=J_y\sin\beta_i+J_z\cos\beta_i$. 
%
The form factors for $\Gamma=x^2-y^2$ and $xy$ are given by multiplying $-1$ to the r.h.s. of each one shown above. 
For $\Gamma=yz$ and $zx$, we have 
$\langle i\ell\pm|\hat{O}_{yz}|i\ell\pm\rangle=
\sqrt{3}
a_{\ell}(1-2a_{\ell})\cos\alpha_{i}\cos\beta_{i}\sin\beta_{i}
/2
$ and 
$\langle i\ell\pm|\hat{O}_{zx}|i\ell\pm\rangle=
\sqrt{3}
a_{\ell}(-1+2a_{\ell})\sin\alpha_{i}\sin\beta_{i}\cos\beta_{i}
/2
$. 
The other ones are the same as above. 

\subsection{Elastic constants}

The elastic constant is given by 
\begin{eqnarray}
C_{\Gamma}=C_{\Gamma}^{(0)}-g_{\Gamma}^2\chi_{\Gamma}, 
\label{eq:elastic_C}
\end{eqnarray}
where $C_{\Gamma}^{(0)}$ is the elastic constant of the background and $g_{\Gamma}$ is the quadrupole-strain coupling constant. 
Here, $\chi_{\Gamma}$ is defined by 
$\chi_{\Gamma}=\lim_{{\bm q}\to{\bm 0}}\chi_{\Gamma}^{\rm F}({\bm q},\omega=0)$ 
with $\chi^{\rm F}_{\Gamma}({\bm q},\omega)=\sum_{ii'}\chi^{ii'}_{O_{\Gamma}}({\bm q},\omega)$. 
Here, 
$C_{B}=(C_{11}+2C_{12})/3$ 
is elastic constant for symmetery strain $\varepsilon_{xx}+\varepsilon_{yy}+\varepsilon_{zz}$, 
$C_v=(C_{11}-C_{12})/2$ for $(2\varepsilon_{zz}-\varepsilon_{xx}-\varepsilon_{yy})/\sqrt{3}$ and $\varepsilon_{xx}-\varepsilon_{yy}$, and 
$C_{44}$ for $\varepsilon_{xy}$, $\varepsilon_{yz}$, and $\varepsilon_{zx}$~\cite{Luthi}. 

Figure~\ref{fig:chi}(b) shows the temperature dependence of 
$
a_{\rm B}^4e^2/
\chi_{\Gamma}$ 
at the QCP of the valence transition identified by the RPA, i.e., for $(\varepsilon_{f}^{RPA}, U_{fd}^{RPA})$. 
Here, $a_{\rm B}$ is the Bohr radius. 
Below $T=1.0\times 10^{-4}$, which is estimated to be 4.7~K if we assume $(pp\sigma)\approx 4.01$~eV as above, the enhancement of $\chi_{\Gamma}$ becomes prominent 
as seen in $\Gamma=x^2+y^2+z^2$. 

Since the positive sign and negative sign appear in the 4f and 5d terms respectively in Eq.~(\ref{eq:Oi}) for $\Gamma=x^2+y^2+z^2$, $\chi_{x^2+y^2+z^2}$ is classified into the relative-charge fluctuation. This is confirmed by the fact that the $T$ dependence of $1/\chi_{x^2+y^2+z^2}$ is well scaled by $1/\chi_{n_f-n_d}^{\rm F}$, i.e., $a_{\rm B}^4e^2/\chi_{x^2+y^2+z^2}\approx 6.5e^2/\chi_{n_f-n_d}^{\rm F}$, as shown in Fig.~\ref{fig:chi}(b).

On the other hand, since the eigenvalues for the electric quadrupole modes for $\Gamma=2z^2-x^2-y^2$, $x^2-y^2$, $xy$, $yz$, and $zx$ do not appaoach 1 for $T\to 0$ in Fig.~\ref{fig:chi}(a) it turns out that the quadrupole fluctuations $\chi_{\Gamma}$ do not diverge at $T=0$. However, the enhancement in $\chi_{\Gamma}$ is induced around the lowest temperatures by enhanced charge fluctuations mainly due to enhanced $\chi^{\rm F}_{n_f}$ arising from the valence QCP. 
Here, $\chi^{\rm F}_{n_f (n_d)}$ is obtained by setting only the 4f (5d) term in $\chi_{n_f\pm n_d}^{\rm F}$ defined above [see Eq.~(\ref {eq:chi_nf_nd1})]. As shown in Fig.~\ref{fig:nfEf}(b), $\chi_{n_f}^{\rm F}$ is indeed  much larger than $\chi_{n_d}^{\rm F}$. Thus 
it is understandable from the fact that $\chi_{n_f}^{ii'}({\bm q},\omega)$ contributes dominantly to the r.h.s. of Eq.~(\ref{eq:chi_E4_2}).

Although all the modes show the tendency of increase in $\chi_{\Gamma}$ for $T\to 0$, the enhancement is less weak for $\Gamma=x^2-y^2$ and $xy$, as shown in Fig.~\ref{fig:chi}(b). This is due to each value of the corresponding form factors (see Appendix) where the off-diagonal component with respect to the Yb site in $\chi_{\Gamma}$, i.e. $\lim_{{\bm q}\to {\bf 0}}\chi^{ii'}_{O_{\Gamma}}({\bm q},\omega=0)$ for $i\ne i'$, are cancelled partly for $\Gamma=x^2-y^2$ by their plus and minus contributions and is diminished mostly for $\Gamma=xy$ by vanishment of the form factors except for $i=1$, 3, 11, and 12. 

This indicates that softening in the bulk modulus $C_{B}$ occurs most remarkably and also in $C_v$ at low temperatures for $T\ll\Delta$. 
Interestingly, in the high-$T$ side of Fig.~\ref{fig:chi}(b), the enhancement of $\chi_{\Gamma}$ also occurs as remarkably seen for $\Gamma=x^2+y^2+z^2$, which is also due to the form factors (see Appendix) where the off-diagonal component $\lim_{{\bm q}\to {\bf 0}}\chi^{ii'}_{O_{\Gamma}}({\bm q},\omega=0)$ for $i\ne i'$ gives rise to the reduction of $\chi_{\Gamma}$ around $T=2.0\times 10^{-4}$. 

It is noted that in the polycrystal the vanishment of the form factors for $\Gamma=xy$ 
may 
become incomplete since the quantization axis of the CEF at each Yb site deviates from the one shown in Fig.~\ref{fig:YbAlAu}(f) 
in the different domains, 
giving rise to finite values of $\langle i\ell\pm|\hat{O}_{\Gamma}|i\ell\pm\rangle$. 
This effect is expected to cause the 
tendency of 
softening even in the transverse mode $\Gamma=xy$, as shown in the longitudinal mode $\Gamma={x^2}$ in Fig.~\ref{fig:chi}(b). 
This also applies to the QC. 
We note that 
this conclusion does not depend on the detail of the orientation of the $\Psi^{4f}(\hat{\bm r}_i)$ in the CEF ground state [see Fig.~\ref{fig:YbAlAu}(f)]~\cite{Sugimoto,Hiroto2020} since the deviation of the $xyz$ coordinate at each Yb site in the domains in the polycrystal is essential for causing the softening in both modes. 


\section{Conclusion and discussion}

We have constructed the realistic effective model for the AC Yb$_{14}$Al$_{35}$Au$_{51}$. The Coulomb repulsion between the 4f and 5d electrons at Yb is evaluated as $U_{fd}\approx 1.46$~eV for realizing the QCP of Yb-valence transition. The RPA analysis of the QCP has shown that softening in elastic constants occurs prominently for bulk modulus and longitudinal mode at low temperatures.

At low temperatures shown in Fig.~\ref{fig:chi}(b), the contributions from the lattice are expected to be smaller than the contributions from electrons. 
Since the CVF dominate over other contributions such as phonon and phason at low temperatures shown in Fig.~\ref{fig:nfEf}(b), it is expected that the softening occurs in the QC as well as the pressure-tuned AC. This is because the deviation of the local $xyz$ coordinate at each Yb site from laboratory coordinate $x'y'z'$ (see Fig.~\ref{fig:YbAlAu}) is essential for the emergence of the softening, which is commonly realized in both the QC and AC.

When pressure is applied to the AC Yb$_{14}$Al$_{35}$Au$_{51}$ and is tuned to $P=1.96$~GPa, the QCP appears~\cite{Matsukawa}. 
There the softening in the elastic constants as noted above is expected to be observed. 
In the QC, the recent measurement has revealed that the QCP of the Yb transition is realized~\cite{NKSato2019}. Hence, the softening in the elastic constants is expected to appear for $T\ll\Delta$. Examination of our results by ultrasonic measurements in the AC under pressure as well as the QC are interesting as the future subject.

\section{Acknowledgments}
The author thanks N. K. Sato, T. Ishimasa,  K. Deguchi, and K. Imura for valuable discussions. 
This work was supported by JSPS KAKENHI Grant Numbers JP18K03542, JP18H04326, and JP19H00648. 

\section{Appendix: Mechanism of the temperature dependences of $\chi_{\Gamma}$}

In this {Appendix, 
the mechanism of the temperature dependences of $\chi_{\Gamma}$ for $\Gamma=x^2-y^2$, $2z^2-x^2-y^2$, $xy$, $x^2+y^2+z^2$, and $x^2$ shown in 
Fig.~\ref{fig:chi}(b) 
is explained.

The explicit form of $\langle i\ell\pm|\hat{O}_{x^2}|i\ell\pm\rangle$ 
in Eq.~(\ref{eq:Omat})  
is given by 
\begin{eqnarray}
\langle i\ell\pm|\hat{O}_{x^2}|i\ell\pm\rangle=\frac{a_{\ell}}{2}(\cos^2\alpha_i\cos^2\beta_i+\sin^2\alpha_i)+a_{\ell}^2\cos^2\alpha_i\sin^2\beta_i.
\label{eq:Ox2}
\end{eqnarray}
In Table~\ref{tb:Oi}, the 
expectation values of $\langle i1\pm|\hat{O}_{\Gamma}|i1\pm\rangle$ 
on the 4f state 
in Eq.~(\ref{eq:Omat}) for $\Gamma=x^2-y^2$, $2z^2-x^2-y^2$, $xy$, $x^2+y^2+z^2$, and $x^2$ at the $i$th Yb site are listed.

%
\begin{table}[h]
\begin{center}
\begin{tabular}{crrrcc}
\hline
\multicolumn{1}{c}{$i$} &\multicolumn{1}{c}{$x^2-y^2$} & \multicolumn{1}{c}{$2z^2-x^2-y^2$} & 
\multicolumn{1}{c}{\hspace*{0.3cm}$xy$} &\multicolumn{1}{c}{$x^2+y^2+z^2$} & \multicolumn{1}{c}{$x^2$} 
\\
\hline
$1$ & ${-2.00}$ & ${-21/8}$ & \hspace*{0.3cm} ${8.17}$ & $63/4$ & $4.69$ \\
$2$ & ${-1.27}$ & ${3.04}$ & \hspace*{0.3cm}$0$ & $63/4$ & $4.69$ \\
$3$ & ${-2.00}$ & ${-21/8}$ & \hspace*{0.3cm}${-8.17}$ & $63/4$ & $4.69$ \\
$4$ & ${-1.27}$ & ${3.04}$ & \hspace*{0.3cm}$0$ & $63/4$ & $4.69$ \\
$5$ & ${3.27}$ & ${-0.42}$ & \hspace*{0.3cm}$0$ & $63/4$ & $1.75$ \\
$6$ & ${3.27}$ & ${-0.42}$ & \hspace*{0.3cm}$0$ & $63/4$ & $1.75$ \\
$7$ & ${-1.27}$ & ${3.04}$ & \hspace*{0.3cm} $0$ & $63/4$ & $4.69$ \\
$8$ & ${3.27}$ & ${-0.42}$ & \hspace*{0.3cm}$0$ & $63/4$ & $1.75$ \\
$9$ & ${3.27}$ & ${-0.42}$ & \hspace*{0.3cm}$0$ & $63/4$ & $1.75$ \\
$10$ & ${-1.27}$ & ${3.04}$ & \hspace*{0.3cm}$0$ & $63/4$ & $4.69$ \\
$11$ & ${-2.00}$ & ${-21/8}$ & \hspace*{0.3cm}${-8.17}$ & $63/4$ & $4.69$ \\
$12$ & ${-2.00}$ & ${-21/8}$ & \hspace*{0.3cm}${8.17}$ & $63/4$ & $4.69$ \\
\hline
\end{tabular}
\end{center}
\caption{$\langle i1\pm|\hat{O}_{\Gamma}|i1\pm\rangle$ for $\Gamma=x^2-y^2$, $2z^2-x^2-y^2$, $xy$, $x^2+y^2+z^2$, and $x^2$ at the $i$th Yb site.}
\label{tb:Oi}
\end{table}

In Table~\ref{tb:RPA_sus}, the values of the RPA susceptibility of the 4f orbital with the Kramers state $\eta=+$, i.e. $\lim_{{\bm q}\to{\bf 0}}\chi^{ii'}_{1+1+1+1+}({\bm q},\omega=0)$ [see Eq.~(\ref{eq:chi_def})], are listed for the $i=1$ and $i'=1$--$12$ at $T=1.0\times 10^{-5}$, $3.0\times 10^{-5}$, and $2.0\times 10^{-4}$ as a representative case. 
The diagonal component $(i,i')=(1,1)$ has a positive value, which increases sharply on cooling below $T=3.0\times 10^{-5}$. 
On the other hand, in the off-diagonal component for $i\ne i'$, negative values appear as seen in Table~\ref{tb:RPA_sus} at $T=2.0\times 10^{-4}$. As $T$ decreases, these negative values change to positive values finally as seen at $T=1.0\times 10^{-5}$.

%
\begin{table}[h]
\begin{center}
\begin{tabular}{ccrrr}
\hline
\multicolumn{1}{c}{$i$} &\multicolumn{1}{c}{$i'$} & \multicolumn{1}{r}{$T=1.0\times 10^{-5}$} & 
\multicolumn{1}{r}{$3.0\times 10^{-5}$} &\multicolumn{1}{r}{$2.0\times 10^{-4}$} 
\\
\hline
$1$ & $1$ & $27.10$ & $21.74$ & $21.08$ \\
$1$ & $2$ & $3.41$ & $-0.70$ & $-0.99$ \\
$1$ & $3$ & $7.08$ & $3.07$ & $2.84$ \\
$1$ & $4$ & $3.41$ & $-0.70$ & $-0.99$ \\
$1$ & $5$ & $3.41$ & $-0.70$ & $-0.99$ \\
$1$ & $6$ & $3.41$ & $-0.70$ & $-0.99$ \\
$1$ & $7$ & $2.78$ & $-0.99$ & $-1.17$ \\
$1$ & $8$ & $2.78$ & $-0.99$ & $-1.17$ \\
$1$ & $9$ & $2.78$ & $-0.99$ & $-1.17$ \\
$1$ & $10$ & $2.78$ & $-0.99$ & $-1.17$ \\
$1$ & $11$ & $5.22$ & $0.37$ & $-0.15$ \\
$1$ & $12$ & $3.94$ & $0.31$ & $0.19$ \\
\hline
\end{tabular}
\end{center}
\caption{RPA susceptibility $\lim_{{\bm q}\to{\bf 0}}\chi^{ii'}_{1+1+1+1+}({\bm q},\omega=0)$ at $T=1.0\times 10^{-5}$, $3.0\times 10^{-5}$, and $2.0\times 10^{-4}$.}
\label{tb:RPA_sus}
\end{table}

The electric quadrupole susceptibility is calculated in Eq.~(\ref{eq:chi_E4_2}), which is reflected in the elastic constants as discussed in the main text. 
Since 
$\langle i1\pm|\hat{O}_{x^2+y^2+z^2}|i1\pm\rangle$
 are all positive at the $i$th Yb site, the above negative-to-positive change in the off-diagonal components all contribute to increase in $\chi_{x^2+y^2+z^2}$ on cooling from $T=2.0\times 10^{-4}$ to $T=1.0\times 10^{-5}$. 
This results in the convex curve of the $a_{\rm B}^4e^2/\chi_{x^2+y^2+z^2}$ vs. $T$ plot in Fig.~\ref{fig:chi}(b). 
The same applies to $\chi_{x^2}$ since 
$\langle i1\pm|\hat{O}_{x^2}|i1\pm\rangle$ are all positive, as shown in Table~\ref{tb:Oi}. 

For $\Gamma=xy$, the form factor is zero at the $i=2$, 4, 5, 6, 7, 8, 9, 10th Yb site, as shown in Table~\ref{tb:Oi}. 
This implies that the above negative-to-positive change in the off-diagonal component in the RPA susceptibility (see Table~\ref{tb:RPA_sus}) does not contribute to increase in $\chi_{xy}$, which is in sharp contrast to $\chi_{x^2+y^2+z^2}$ and $\chi_{x^2}$. Hence, the main contribution comes form the diagonal component for $i=i'$ in Table~\ref{tb:RPA_sus}, which results in almost flat-$T$ dependence of $\chi_{xy}$ 
in Fig.~\ref{fig:chi}(b). 

For $\Gamma=x^2-y^2$ and $2z^2-x^2-y^2$, the form factors are not zero but have positive and negative values depending on the $i$th Yb site, as shown in Table~\ref{tb:Oi}. 
Hence, the off-diagonal components in the RPA susceptibility are cancelled partly as $T$ decreases, which results in the gradual increase in $\chi_{\Gamma}$ around the lowest temperatures 
in Fig.~\ref{fig:chi}(b).



\begin{thebibliography}{99}
\bibitem{noteCEF} This is a literal interpretation based on the point-charge model. In reality, screening effect by conduction electrons is expected to reduce the charge of the ligand ion in the metallic crystal.
\bibitem{note_C} Note that at the valence QCP, the CVF beyond the MF theory contribute to the mass enhancement, giving rise to the enhanced specific-heat coefficient. See, for detail, ref.~\cite{WMssc}.
\end{thebibliography}


\end{document}